\documentclass[conference]{IEEEtran}

\IEEEoverridecommandlockouts
\usepackage{cite}
\usepackage{amsfonts}
\usepackage{textcomp}
    
\usepackage{graphicx}
\usepackage{balance}  

\def\BibTeX{{\rm B\kern-.05em{\sc i\kern-.025em b}\kern-.08em
    T\kern-.1667em\lower.7ex\hbox{E}\kern-.125emX}}
\usepackage{booktabs} 

\usepackage{times}
\usepackage{url}
\usepackage{amsmath}
\usepackage{listings}
\usepackage[noend]{algpseudocode}
\makeatletter
\def\algbackskip{\hskip-\ALG@thistlm}
\let\OldStatex\Statex
\renewcommand{\Statex}[1][3]{%
  \setlength\@tempdima{\algorithmicindent}%
  \OldStatex\hskip\dimexpr#1\@tempdima\relax}
\makeatother
\usepackage{tikz}
\usepackage{amssymb}
\usepackage{pifont}
\newcommand{\cmark}{\ding{51}}%
\newcommand{\xmark}{\ding{55}}%
\usepackage{xcolor}

\colorlet{punct}{red!60!black}
\definecolor{delim}{RGB}{20,105,176}
\colorlet{numb}{magenta!60!black}

\lstdefinelanguage{json}{
    basicstyle=\normalfont\ttfamily,
    numbers=left,
    numberstyle=\scriptsize,
    stepnumber=1,
    numbersep=4pt,
    showstringspaces=false,
    breaklines=true,
    frame=lines,
    literate=
     *{0}{{{\color{numb}0}}}{1}
      {1}{{{\color{numb}1}}}{1}
      {2}{{{\color{numb}2}}}{1}
      {3}{{{\color{numb}3}}}{1}
      {4}{{{\color{numb}4}}}{1}
      {5}{{{\color{numb}5}}}{1}
      {6}{{{\color{numb}6}}}{1}
      {7}{{{\color{numb}7}}}{1}
      {8}{{{\color{numb}8}}}{1}
      {9}{{{\color{numb}9}}}{1}
      {:}{{{\color{punct}{:}}}}{1}
      {,}{{{\color{punct}{,}}}}{1}
      {\{}{{{\color{delim}{\{}}}}{1}
      {\}}{{{\color{delim}{\}}}}}{1}
      {[}{{{\color{delim}{[}}}}{1}
      {]}{{{\color{delim}{]}}}}{1},
}

\begin{document}
\sloppy


\title{Consentio: Managing Consent to Data Access using Permissioned Blockchains}







\author{\IEEEauthorblockN{ Rishav Raj Agarwal*, Dhruv Kumar*\thanks{\textsuperscript{*}Equal contribution by both authors.}, Lukasz Golab, Srinivasan Keshav}
\IEEEauthorblockA{
\textit{University of Waterloo, Waterloo, Canada}\\
{\{rragarwal,d35kumar,lgolab,keshav\}@uwaterloo.ca}}


}

\maketitle

\begin{abstract}
The increasing amount of personal data is raising serious issues in the context of privacy, security, and data ownership. 
Entities whose data are being collected can benefit from mechanisms to manage the parties that \textit{can} access their data and to audit who \textit{has} accessed their data. 
Consent management systems address these issues.
We present \textit{Consentio}, a scalable consent management system based on the Hyperledger Fabric permissioned blockchain. 
The challenge we address is to ensure high throughput and low latency of endorsing data access requests and granting or revoking consent. 
Experimental results show that our system can handle as many as 6,000 access requests per second, allowing it to scale to very large deployments.
\end{abstract}

\section{Introduction}
\label{intro}
Every day 3.7 billion individuals access the Internet, generating 2.5 quintillion bytes of data\footnote{https://www.forbes.com/sites/bernardmarr/2018/05/21/how-much-data-do-we-create-every-day-the-mind-blowing-stats-everyone-should-read}.
The increasing amount of data being collected, especially personal data, is raising serious issues in the context of privacy, security, and data ownership.
These issues have been recognized, for example, by the recent EU General Data Protection Regulation (GDPR), 
which requires organizations consuming private data to obtain consent from the individuals whose data are being collected.

We illustrate the broad need for consent in accessing private data in three sensitive contexts.
First, in healthcare, personal health data such as medical histories, vital signs, and lab test results are collected by hospitals, wearables, health-tracking applications, and assisted living systems.
The data may be shared with healthcare professionals to provide care or to participate in research studies~\cite{tang2006personal}. 
However, these entities should only be able to view personal data with a patient's explicit consent.
Second, a great deal of personal data are collected by online and mobile apps, and often shared with third parties: emails, web browsing and search histories, location data.
Recent developments such as the Cambridge Analytica scandal~\cite{isaak2018user} show that users have little control over how their online data are used or shared. Furthermore, data owners are often unable to determine which parties have access to the data, and there have been cases where companies have shared personal data (for instance, Skype recordings\footnote{https://techcrunch.com/2019/08/15/microsoft-tweaks-privacy-policy-to-admit-humans-can-listen-to-skype-translator-and-cortana-audio/}) with third parties without explicit permission from the data owners.  
A data-access
process that includes user consent could have averted or at least
minimized this privacy breach.
Finally, smart cities are collecting potentially sensitive data through street cameras and smart electricity meters.
For example, smart meter data are sent to utility companies for billing, but individuals may want to share their smart meter data with analytics services to help reduce their bills, motivating energy data sharing platforms such as Green Button\footnote{\url{http://www.greenbuttondata.org/}}.
Without a consent management mechanism, users may not be willing to upload their data to such platforms. 

Ideally, individuals whose data are being collected should be able to decide who \emph{can} access their personal data and audit who \emph{has} accessed their personal data with the help of a \emph{consent management system} (CMS). 
A CMS acts as a bridge between skeptical data owners and data consumers by adding access control and transparency. 
However, when a single entity, such as Facebook, controls the CMS, data owners are forced to trust this entity, even if they would not
wish to do so. We believe that data owners would be more willing to trust a consortium of non-colluding
entities over a single entity, just as citizens of a democracy trust Parliament, but not necessarily every individual Member of Parliament.
This intuition naturally suggests using blockchains to manage consent in a decentralized and verifiable manner, as has been done in prior work~\cite{zyskind2015decentralizing,dias2018blockchain,zhang2018smart,ekblaw2016case}.  
In this approach, a blockchain transaction corresponds to an individual's granting or withdrawing of consent to share data with a particular third party, or a third party's request to access data.
Individuals submit transactions to specify who can access their data and can audit the blockchain to find out who was granted access to their data.
The blockchain is tamper-evident, meaning that consent cannot be forged. Moreover, having parties from different organizations approving transactions removes the requirement to trust a single entity.

Unfortunately, most prior work in this area is not only domain-specific but also lacks implementation details.
As a result, it remains unclear if a blockchain can be the basis for a \textit{usable} CMS at scale.
To fill this gap, this paper makes the following contributions.
\begin{enumerate}
    \item \textbf{CMS design:} We present \emph{Consentio}, a general and scalable CMS with a blockchain back end, mapping consent operations to blockchain transactions.
    \item \textbf{CMS implementation:} We implement \emph{Consentio} using Hyperledger Fabric~\cite{androulaki2018hyperledger}, a state-of-the-art permissioned blockchain. 
    Our solution is implemented in Hyperledger Fabric using \emph{smart contracts}.
   
    \item \textbf{CMS world state design:} Fabric maintains a \emph{world state} key-value store to speed up transaction proessing. 
    It is not obvious what world state should mean in consent management. We address this challenge by (a) examining the space of possible key-value world state designs for consent management and (b) proposing a design that ensures high throughput and low latency of endorsing and revoking consent.   Experimental results using a Fabric cluster show that \emph{Consentio} can handle as many as 6,000 access requests/s and scales well with the number of nodes used to endorse transactions.
\end{enumerate}

We believe that a novel aspect of our solution is that it ``does more with less.'' 
Hyperledger Fabric uses a simple key-value store to store world state, 
rather than relational database system that supports SQL.   
Yet, we demonstrate that an efficient and scalable CMS can be built using Fabric.

It is important to note that, similar to any other CMS, \emph{Consentio} only
manages and audits \textit{consent} to data access rather than data access itself.  
That is, when \emph{Consentio} approves a request to access data, it effectively gives the requesting entity a key to obtain the data from 
a trusted data store, while maintaining a record of this transaction. This assumes that data stores are trusted to release data only after ensuring that access is permitted, and do not have `back doors.' This is not an onerous assumption, in that it is widely supported by services such as Dropbox, Microsoft's OneDrive, and Google Drive. 
Any data store that leaks data without correctly checking permissions can face legal action and therefore has no incentive to 
release data to unauthorized users. 
Thus, we focus on the design of a tamper-evident and decentralized CMS, something that existing data storage solutions do not currently
provide. 
Furthermore, after obtaining the desired data, the requesting entity is trusted not to share the data with unauthorized parties.  
Again, we do not address this problem, though we note that it can
be solved using mechanisms such as data watermarking~\cite{podilchuk2001digital}. 


\section{Consent Management: Definitions and Goals}
\label{approach}

In this section, we discuss the participating entities, functionality, and example use cases of a Consent Management System. 
We begin with some definitions based on the consent management and role-based access control literature \cite{bertino2001trbac,sandhu1996role}.

\begin{itemize}
 
\item An \textbf{Individual}, also referred to as a Data Subject, is a person whose data are being collected.

\item A \textbf{Resource} is a subset of an individual's data of the same type, e.g., a time series of blood pressure measurements.

\item A \textbf{Data Generator} is an entity that produces data, e.g., a hospital or a smart meter system, and stores this data in a \textbf{Data Store}. 

\item A \textbf{Data Consumer}, also referred to as a Data Processor, is an entity that may request access to private data (from the data store) to perform  \textbf{Data Analysis}. 

\item Data consumers are grouped into \textbf{Roles}; a given data consumer may be assigned multiple roles.  For example, a doctor at hospital X may be assigned the roles of ``Cancer researcher'' and/or ``Hospital X staff''. 

\item A \textbf{Watchdog} assigs roles to data consumers. For example, if a doctor wishes to be given a role of ``Cancer researcher'', he or she must contact an appropriate watchdog (e.g., a research ethics board) for approval.

\item \textbf{Consent} is expressed by a set of rules that specify which role(s) may access an individual's resources.


\end{itemize}

Next, we describe the required CMS functionality, drawing again from the relevant consent management literature. Note that here the CMS acts as the Data Controller. 

\begin{enumerate}


\item Individuals grant or withdraw consent to data consumers acting in particular roles. Consent specification is \emph{fine-grained}: it may depend on the resource, the role, or the time range of the data (see, e.g.,~\cite{dias2018blockchain,ekblaw2016case,zyskind2015decentralizing}). In practice, some individuals, such as infants, may not be able to give consent, motivating the need to \emph{delegate} consent to another individual. 

\item Watchdogs assign and revoke data consumers' roles.

\item Given their roles, data consumers request permission to access data (see, e.g.,~\cite{zyskind2015decentralizing,hashemi2017decentralized}).  The CMS must determine whether such permission can be granted.

\item A CMS must be \emph{tamper-proof} and \emph{auditable}.  Individuals may audit who requested (and was granted) access to their data (see, e.g.,~\cite{hashemi2016world,zyskind2015decentralizing}). Data consumers may audit the CMS if asked to prove that they have obtained clearance to access data.

\end{enumerate}

Below, we describe the
entities participating in consent management in three exemplary use cases: 

\begin{itemize}
  \item  \textbf{Electronic Health Records}: Here, patients correspond to individuals.  Hospitals and healthcare research organizations generate data using medical equipment and store the data in a hospital database.  Doctors and other healthcare staff may be the data consumers; they may require access to personal data for treatment or to carry out research studies. Roles may be assigned to consumers based on their affiliations or occupations.  Watchdogs such as research ethics boards may approve or deny roles such as those required to do medical research.  Patients may audit the CMS to determine who requested access to their data, while doctors and medical researchers may use the CMS to prove that they requested and obtained permission to access the data. 

    \item  \textbf{Smart Infrastructure}: With the growth of IoT technologies, governments are setting up smart infrastructure in urban areas. Citizens use public infrastructure, and should know if a third party has been given access to their data, and, if possible, should control access to the data.  For instance, utility companies may receive smart meter data for billing.  Homeowners may want to share smart meter data with third parties or academic researchers to help reduce their bills. These third parties may be verified by civil society organizations before being granted a role of, e.g., an energy data analyst.  Individuals may audit the CMS to determine who accessed their data and third parties may query the CMS to produce proof that they obtained consent to access the data.

    \item  \textbf{Social Media}: Social media companies may sell user data to third parties or use the data to serve advertisements. Users should be able to see who accessed their data and manage their consent settings. Individuals are the people using social media applications, and the data may be stored on social media company databases.  Company analysts and third parties act as data consumers, and they should only be able to access data after receiving consent from users through the CMS. Privacy advocates serve as watchdogs here.  
\end{itemize}

\section{Blockchains for Consent Management}
\label{trust}

In this section,
we justify the use of a permissioned blockchain for consent management. We also
give a high-level overview of \emph{Consentio}'s design, scope and limitations. Implementation details follow in Section~\ref{implementation}.


First, assume a centralized architecture for private data management, in which the CMS, the data generators and the data store are controlled by the same centralized entity.
In this design, individuals, roles and watchdogs all rely on the controlling entity to give or revoke consent, assign and revoke roles, request access to data, and audit data request histories.
As a result, the following trust relationships exist:

\begin{enumerate}
    \item Individuals trust the CMS to keep track of their consent settings, and to only allow authorized roles to obtain access to private data.
    \item Individuals trust the watchdogs to assign and revoke roles to data consumers.
    \item Data consumers (assigned particular roles) trust the CMS to correctly allow or deny access to data.
    \item Watchdogs trust the CMS to assign and revoke data consumers' roles as requested.
    \item Data Stores must be trusted to not divulge data to parties who do not have proper consent.
    \item Individuals trust the CMS not to collude with data consumers to endorse data access without the individuals' consent.
    \item Watchdogs trust the CMS not to collude with data consumers to endorse data access to unauthorized roles.
    \item Data consumers must be trusted not to share any data they obtained with unauthorized parties, and to use the data only for approved purposes.
\end{enumerate}

Following recent work in consent management, we observe that blockchains can mitigate some of the trust issues identified earlier due to the presence of a centralized third party that runs the CMS.  
Furthermore, as also observed in previous work (as discussed in Section~\ref{relatedwork}), we choose a permissioned blockchain in order to control membership.
Permissioned systems are usually jointly owned by a consortium of known participants who may not necessarily trust each other.
Here, nodes are not anonymous, and each node must be approved to join the network by a \emph{membership service} run by the consortium. Thus, if any actor is found to engage in malicious activities, the membership service can take appropriate action.

\begin{figure*}[t]
\centering
 \includegraphics[width=0.7\linewidth]{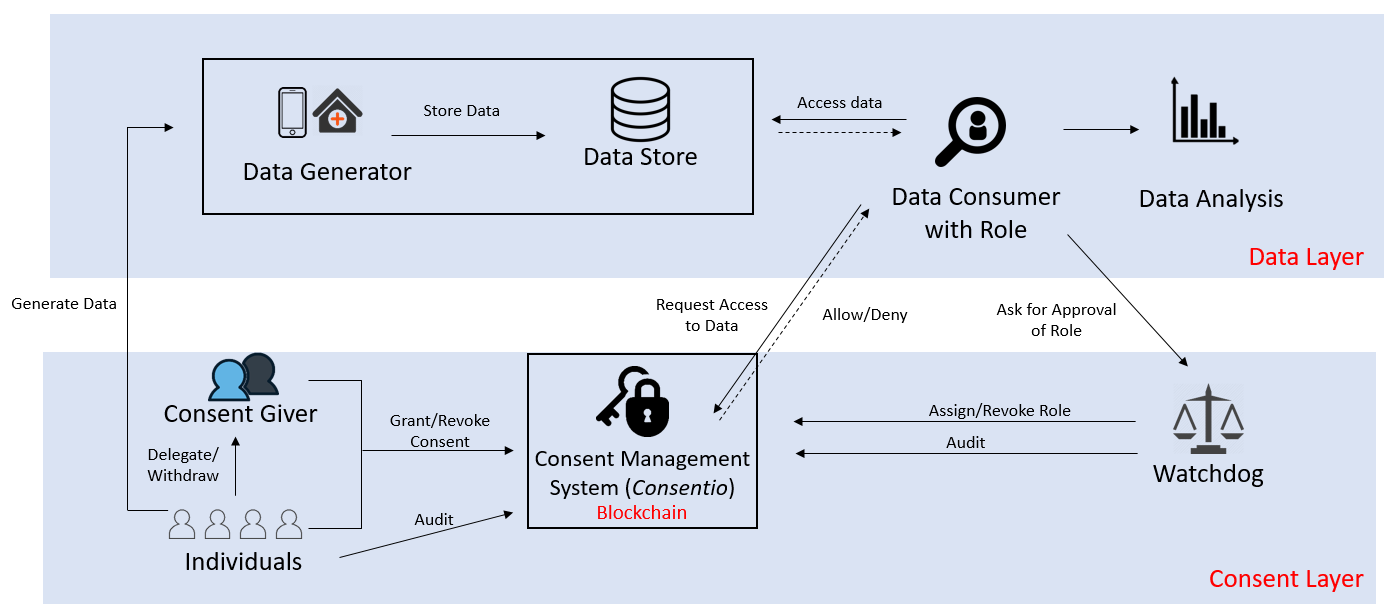}
  \caption{Consent management using \emph{Consentio}.}
  \label{arch}
\end{figure*}

A critical design element in our solution is to decouple consent management from data management, akin to decoupling the control plane from the data plane in computer networks.  
Figure~\ref{arch} places \emph{Consentio} in this decoupled environment.  
We define the \emph{consent layer}, which is the focus of this paper, to include CMS functionality such as granting or withdrawing consent to personal data.  
In contrast, the \emph{data layer} includes the data themselves.
As shown in the figure, consent is jointly managed by all parties using \emph{Consentio} with a blockchain back end, without having to be entrusted to a third party.
Individuals can submit transactions to give or withdraw consent for a given role to access a given data resource.
Data consumers can request role approval from watchdogs, who can submit the corresponding transactions (to grant or revoke a role for a given consumer).
Furthermore, data consumers acting under some role may submit transactions to request access to data.
If permission is granted, data consumers may take their proof of consent to the data store to obtain the data (the details of obtaining the data given a proof of consent are orthogonal to this paper and not discussed further).
All of these transactions are on the blockchain and therefore are 
auditable by all parties.

Given the design in Figure~\ref{arch}, the following trust issues are mitigated.

\begin{itemize}
    \item The CMS cannot collude with data consumers to endorse unauthorized data access.  Any requests for data access (if granted) will be on the blockchain  
    and may be audited.
    \item The CMS cannot deny data access to consumers who ought to have access.  Individuals' consent settings will be on the blockchain
    and may be verified.
    \item The CMS cannot collude with data consumers to ignore the roles assigned by watchdogs.  All role assignments and revocations will be on the blockchain 
    and may be audited.
\end{itemize}

These situations correspond to trust issues 1, 3, 4, 6 and 7. However, 
trust issues 2, 5, and 8 remain: watchdogs must be trusted; the data store must be trusted to release data only to approved consumers; and data consumers must be trusted to not ``leak'' data.  
Of these three issues, only the watchdog issue is at the consent layer. Note that watchdogs are meant to be trusted public organizations, such as
governments or consortia, and therefore have some built-in transparency and can be held accountable by the public. Hence,
trust in them is reasonable.
We have already discussed why we believe data stores can be trusted. Finally, techniques such as watermarking can mitigate data leaks by data consumers. Thus, Consentio, when embedded in a holistic system with civil society actors and trusted data stores, can provide a reasonably effective 
solution for consent management. We now turn our attention to details of its implemetation and performance characterization.

\section{Implementing Consentio in Fabric}
\label{implementation}



We now discuss the implementation of \emph{Consentio} using the Hyperledger Fabric~\cite{androulaki2018hyperledger} permissioned blockchain system.
Our CMS design generalizes those presented in recent work on healthcare and IoT consent management (see, e.g.,~\cite{zhang2018smart,rantos2018advocate,azaria2016medrec}). 
We start with an overview of Fabric, followed by an exploration of possible world state designs for consent management, and a justification of our design choice.
The smart contract details and source code are publicly available at \url{ https://github.com/ddhruvkr/Consentio}.

\subsection{Overview of Hyperledger Fabric}
\label{overview}

Hyperledger Fabric is an actively studied permissioned blockchain system~\cite{amiri2019parblockchain,gorenflo2019fastfabric,sharma2019blurring,thakkar2018performance}. There are two important data structures: the blockchain, which is the transaction log, and the world state, which is a key-value store to maintain some application-defined state information.  For concurrency control, the key-value store is versioned, and each key includes a version number of its latest value.  From a transaction processing standpoint, there are two important stages: transaction ordering and transaction execution.  

The nodes in a Fabric cluster are divided into \emph{orderers}, responsible only for transaction ordering, and \emph{peers}, each peer maintaining a copy of the blockchain and the world state.  Some peers are also \emph{endorsers}, which are additionally responsible for endorsing client transactions. 

We summarize Fabric's transaction pipeline as follows.  A client sends its transaction to the endorsers.
Each endorser simulates the execution of the transaction 
in a sandboxed environment, and records the versions of all keys that were read or written.  This \emph{read-write} set is appended to the transaction, along with the endorser's signature, and returned to the client. 
After a client collects the endorsements, it sends its transaction along with the endorsements to the orderers, which agree on the order of transactions and segment them into blocks. Next, blocks are sent to all the peers for validation and commitment.  Each peer serially executes the transactions and updates its world state, incrementing the version numbers of updated keys.  A transaction commits if every key in its read-write set (as computed earlier by the endorsers) still has the same version number. Otherwise, this means that a prior transaction has written a new value to a key touched by the current transaction, and the current transaction aborts.  Finally, each peer appends the new block to its copy of the blockchain. 

\subsection{Consentio Transactions}
\label{Transactions}

Recall from Section~\ref{approach} that a CMS deals with individuals (identified by ind\_id) and their resources (identified by res\_id), data consumers (identified by dc\_id) and their roles (identified by role\_id), as well as watchdogs (identified by wd\_id). To allow fine-grained consent specification, resources are divided into timeframes, with a time unit identified by time\_id.  We now elaborate on the four required CMS functionalities:

\begin{enumerate}
    
    \item \textbf{Grant/Revoke Consent:} An individual (with a given  ind\_id) may grant or withdraw consent to allow consumers (with a particular role\_id granted by a particular wd\_id) to access a fragment of a resource res\_id corresponding to time time\_id. 
    To enable fine-grained consent specification, individuals may allow access to specific temporal fragments of the data, and to specific roles approved by specific watchdogs. 
   While consent can be delegated, we do not record delegation as a transaction on the blockchain due to privacy and legal implications.
   Instead, we offload delegation to the membership service, which is trusted. Alternatively, we can improve security by using a solution similar to ControlChain~\cite{pinno2017controlchain}, which employs a separate blockhain for delegations. In the remainder of this paper, we assume for simplicity that if individual x controls the consent settings of individual y through delegation, then individual x will interact with~\emph{Consentio} as individual y.

    \item \textbf{Assign/Revoke Role:} A watchdog with a particular wd\_id may grant or revoke role role\_id for data consumer dc\_id.
    \item \textbf{Access Request:}  A data consumer with a role role\_id granted by watchdog wd\_id may request access to a fragment of resource res\_id corresponding to time time\_id. The CMS (specifically, Fabric endorsers) must identify the ind\_ids of individuals who have allowed role role\_id granted by wd\_id to have access to the requested fragment of the requested resource. A record of this request will be written to the blockchain, together with the list of the consenting individual IDs. Note that we do not allow consumers to directly request access to a particular individual's data.
    \item \textbf{Audit:} The above transactions must be tamper-proof and auditable. Individuals may audit their consent histories and requests to access their data; consumers may audit their access requests; and watchdogs may audit their role approvals and withdrawals.
\end{enumerate}

Fabric transactions are implemented as \emph{smart contracts} and are recorded in the blockchain. Since Fabric is a permissioned system, all participating individuals, data consumers and watchdogs must be approved by a membership service.
In our setting, endorsers and peers may be managed by a consortium of watchdogs.

Since each Fabric peer maintains a copy of the blockchain, we envision an off-chain data analytics layer supported by the peers that allows  participating parties to audit their transactions. This may be implemented in a tool such as Hyperledger Explorer~\cite{dhillon2017hyperledger}. We can also offload consent delegation to the membership service, which is responsible for assigning public-private key pairs to all peers. Note that only the participating peers have a copy of the blockchain, not the clients, meaning that clients do not have access to other clients' consent policies. Instead, individuals can submit audit requests to the endorser(s), who check which data consumers were given access to the individual's data.  Furthermore, individuals can contact several random endorsers to audit the data.

\subsection{Space of World State Designs}
\label{worldstate}

The key-value world state is a critical data structure maintained by Fabric to process transactions. In simple financial applications, the world state is straightforward: the key is an account ID and the value is the current balance in that account. Smart contracts that move funds from one account to another
must verify that there is enough money in the sender's account, and update the world state accordingly.

In the design of \emph{Consentio}, the technical challenge is to translate the complex requirements on CMS transactions into a simple key-value world state. We want to ensure high transaction throughput to scale to large deployments: many individuals, resources, time units, consumers, etc. Additionally, we want to ensure low latency given Fabric's double-spending prevention that aborts transactions attempting to write to a key that has already been written to by another transaction in the same block. If there are many such conflicting transactions, they must be re-issued in the next block, meaning that it may take many blocks until all such transactions are committed. 

To explore the space of world state designs, we observe that the three main entities are individuals, data (resources), and data consumers (roles); watchdogs also participate in a CMS but there are likely to be fewer watchdogs than individuals and consumers. This suggests three designs, explained below and illustrated in key-value format in Listing~\ref{keyval}.

\begin{itemize}
    \item \textbf{Role-oriented world state} (RoWS) groups similar roles together. A key is a concatenation of resource ID, individual ID, watchdog ID and time ID, and a value is a list of role IDs that were given access to the data specified in the key (i.e., the given time fragment of the given resource of the given individual) and were approved by the given watchdog.
    \item \textbf{Resource-oriented world state} (RWS), initially suggested in~\cite{dubovitskaya2017secure},  groups similar resources together. A key is a concatenation of individual ID, role ID, watchdog ID and time ID, and a value is a list of resource IDs for which consent was given as specified in the key (i.e., the given individual has given consent for the particular temporal fragments of the resources to be accessed by the given roles approved by the given watchdogs).
    \item \textbf{Individual-oriented world state} (IWS) groups similar individuals together. A key is a concatenation of resource ID, role ID, watchdog ID and time ID, and a value is a list of individual IDs giving consent to the data specified in the key (i.e., the given temporal fragment of a the given resource being available to the given role approved by the given watchdog).
\end{itemize}

\begin{lstlisting}[language=json,firstnumber=1 , basicstyle=\small, tab=2,numbersep=3pt,
  tabsize=2,linewidth=.8\linewidth,numbers=left,caption={World state designs in key:value format},captionpos=b, label= keyval]
RoWS
{res_id|ind_id|wd_id|time_id:
        [role_id_1,..., role_id_n]}
------------------------------------
RWS
{ind_id|wd_id|role_id|time_id:
        [res_id_1,..., res_id_n]}
-----------------------------------
IWS
{res_id|wd_id|role_id|time_id:
        [ind_id_1,..., ind_id_n]}
\end{lstlisting}

Independently of these three designs, the world state also needs to record role information. Listing~\ref{keyval_w} shows the world state for role assignments and revocations: a key is a triple (role\_id, dc\_id, wd\_id), and the value indicates whether a role was granted or withdrawn. When processing access request transactions, this will allow us to quickly determine if a given data consumer was assigned a given role by a given watchdog. 

\begin{lstlisting}[language=json,firstnumber=1 , basicstyle=\small, tab=2,numbersep=3pt,
  tabsize=2,linewidth=.9\linewidth,numbers=left, ,caption={Watchdog assigning a role to a data consumer},captionpos=b, label= keyval_w]
{role_id|dc_id|wd_id:'assign'}
\end{lstlisting}

\subsection{Complexity Analysis}
\label{analysis}

We now analyze the computational complexity of transactions using different world state designs.  We show a summary in 
Table~\ref{tab:complexity}. In this section, we assume that an access request refers to one particular timeframe of one particular resource.

\begin{table}[t]
    \small
    \centering
        \caption{Computational complexity of transactions for each world state design (n$=$the number of individuals)}
    \label{tab:complexity}
        \bgroup
    \begin{tabular}{|l |l| l| l|}
    \hline
     \textbf{Functionality} & \textbf{RoWS} & \textbf{RWS   } &\textbf{IWS} \\
        \hline
        \textit{Assign/Revoke Role} & $O(1)$ & $O(1)$ & $O(1)$ \\
        \hline
       \textit{Grant/Revoke Consent }& $O(1)$  & $O(1)$ & $O(1)$ \\
        \hline
        \textit{Access Request} & $O(n)$ & $O(n)$ & $O(1)$ \\
        \hline
    \end{tabular}
\egroup
\end{table}

The complexity of a watchdog's assigning or revoking a role is O(1) independently of the world state design. It suffices to check if a key with the given watchdog id, data consumer id and role id exists; we update its value if it exists and we create it otherwise. 

Using \textit{IWS}, transactions that grant or revoke consent and transactions requesting access both have constant-time complexity. 
If a data consumer (claiming a particular role assigned by a particular watchdog) wants to access one resource for a single timeframe, then we make two key lookups. The first is to determine if the role claimed by the data consumer was assigned by the given watchdog. The second is to retrieve the value (i.e., the list of consenting individual IDs) corresponding to the key with the requested resource and timeframe, the consumer's role\_id, and the watchdog that assigned it. 
If an individual wants to grant or revoke consent (for one individual, one resource, one timeframe and one role), this requires one key lookup.
The lookup is to find the key corresponding to the consent being modified (or create it if it does not exist), and to update the value accordingly. We implement the values (lists of individual IDs) as hashmaps, giving constant-time complexity to add or remove individuals, assuming constant time to serialize and deserialize a hashmap stored as a value in the world state.

Using \textit{RWS} and \textit{RoWS}, granting or revoking consent has constant-time complexity. 
One lookup is required to find the key corresponding to the consent being granted or revoked, and update the value (we implement the values as hashmaps, giving constant-time update cost).
Requesting access in \textit{RWS} and \textit{RoWS} has linear complexity with the number of individuals since individual ID is part of the key. For each individual in the system, we concatenate its individual ID with the parameters contained in the request access transaction, and look up all the corresponding keys. We then return the individual IDs where the corresponding value contained the requesting consumer's role (\textit{RWS}) or the requested resource (\textit{RoWS}). 

Informed by the complexity analysis, we select \textit{IWS} for \emph{Consentio}. Furthermore, we expect the number of individuals in a CMS to be higher than the number of resources (e.g., in healthcare, there may be hundreds of thousands of patients but perhaps hundreds of resources). Thus, the number of keys in \textit{IWS} is expected to be smaller than in the other two designs. 



\subsection{Latency Implications}

Due to Fabric's double spending prevention, transactions accessing the same key (with at least one transaction writing a new value) are said to conflict, and only one of them can be committed in a single block. 
This increases latency since the conflicting transactions must be re-issued and re-endorsed in subsequent blocks.  
If there are $k$ transactions in one block accessing the same key, it will take $k$ blocks to commit all of them, one per block.
Since our three world state designs have different keys, they also induce different transaction collisions.

In \textit{RoWS} and \textit{RWS}, request access transactions are likely to conflict with consent modification transaction because the former are likely to touch many keys, perhaps including the keys being modified by the latter.
This is not the case in \textit{IWS}, where an access request transaction touches only one key. As long as that key is not being modified by another transaction, there is no conflict. Furthermore, regardless of the world state design, access request transactions never conflict with each other because they do not write to the world state (more precisely, they do not modify the version number of any key), only the to blockchain.  

However, consent management transactions for different individuals may conflict with each other only in \textit{IWS} (because in the other two designs, individual ID is part of the key). As a consequence, multiple individuals cannot add or revoke consent for the same resource, same timeframe, same role and same watchdog in a single block.
This may be a problem if multiple individuals wish to revoke consent for a misbehaving role as soon as possible.
However, since roles are approved and withdrawn by watchdogs, it may make more sense for watchdogs to handle such cases. Regardless of the world state design, watchdog transactions do not conflict with any other transactions since they use a separate key space. Thus, watchdogs can immediately revoke a role, and the end result is the same as if all the individuals had revoked all of their consent for that role, one resource-timeframe pair at a time.

\section{Experiments}
\label{experiments}

We evaluate \emph{Consentio} through micro-benchmarking and a comparison of the \textit{IWS} world state design to 
\textit{RWS} (we omit \textit{RoWS}, which, despite the same worst-case complexity as \textit{RWS}, was less efficient due to its larger key space if there are fewer roles than resources).
For the Fabric cluster, we use five local servers connected by a 1 Gbit/s switch. 
Each server is equipped with two Intel Xeon CPU E5-2620 v2 processors at 2.10 GHz, and 64 GB of RAM. 
We use two endorsers, one orderer, one anchor peer, one storage peer, and $100$ clients. The peers are on different servers and the clients are on the same servers as the endorsers.
We use GO to implement  smart contracts and use FastFabric~\cite{gorenflo2019fastfabric} as the underlying permissioned blockchain (FastFabric is a recent modification of Fabric that improves transaction throughput). 

The experiments focus on throughput. We generate non-conflicting transactions that proceed through the entire transaction pipeline without being aborted. Thus, we evaluate the maximum capacity of the pipeline, keeping in mind that conflicting transactions reduce throughput and increase latency.

We do not measure endorsement time since our focus is on speeding up transaction commitment through a suitable world state design.  Additionally, we can easily scale out the endorsers for higher endorsement throughput. Thus, we send pre-endorsed transactions 
to the orderer, which groups them into blocks and sends the blocks to the committer. The committer validates and commits changes to its world state, and sends validated blocks to the endorsers.
We set up 25 threads within each client (totalling 100 threads) that send transactions to the orderer and monitor the time it takes to send 100,000 transactions. Following prior work that has done a detailed micro-benchmarking of FastFabric \cite{gorenflo2019fastfabric}, we set the block size to 100.

Recall from Section~\ref{implementation} that in our context, a data consumer's request to access a resource only reads from the world state: it suffices to read the consent information captured in the world state to decide whether to grant the request.  On the other hand, when an individual modifies their consent settings (or when watchdogs update roles), the world state is also modified.

\subsection{Micro-Benchmarking of IWS World State}

\begin{figure}[t]

 \includegraphics[width=0.77\linewidth]{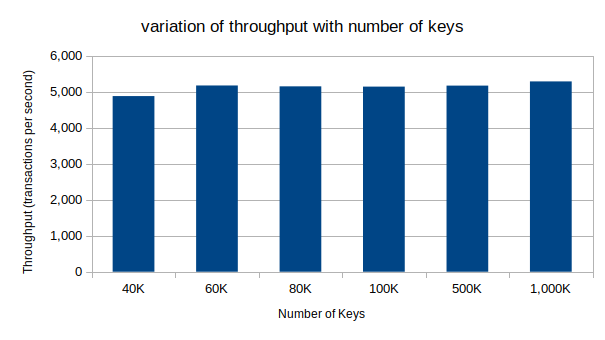}
  \caption{Read throughput performance vs.\ size of key space (the value space and keys touched per transaction are kept constant at 100)}
  \label{exp1}
\end{figure}

\begin{figure}[t]

 \includegraphics[width=0.77\linewidth]{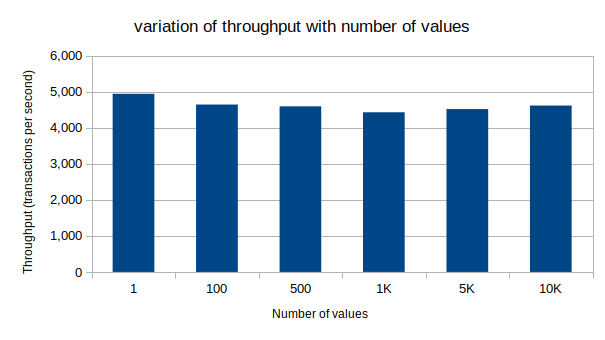}
  \caption{Read throughput performance vs.\ size of value space (keys touched per transaction is kept constant at 100 and key space is kept constant at 20,000)}
  \label{exp3}
\end{figure}

\begin{figure}[t]

 \includegraphics[width=0.77\linewidth]{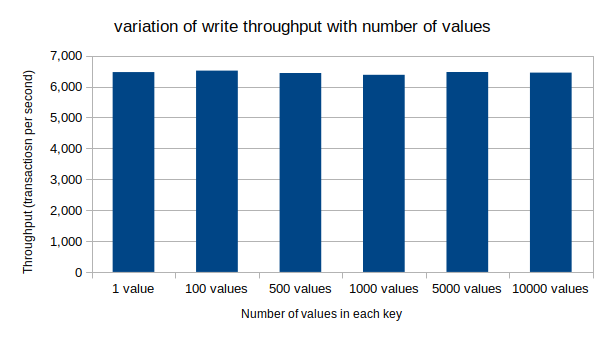}
  \caption{Write throughput performance vs.\ value size per key (key touched per transaction are kept constant at 100 and key space is kept constant at 20,000)}
  \label{exp4}
\end{figure}

\textbf{Effect of world state size:} We vary the number of keys in the world state from 20,000 to 1,000,000, and we set the value size per key (i.e., the number of individual IDs) and keys touched per transaction to 100. We only perform world state reads. Figure~\ref{exp1} shows the transaction throughput. Next, as shown in Figure~\ref{exp3}, we keep the keyspace fixed at 20,000 and keys touched per transaction at 100, and we increase the value space, i.e., the number of individual IDs per key, from 1 to 10,000. Again, we only perform GET requests. Finally, in Figure~\ref{exp4} we show how the \emph{write} throughput (i.e., PUT requests) of the world state changes by increasing the value space. We fix keys touched per transaction at 100 and the key space at 20,000. We choose the number of keys to be 20,000, as this may correspond to parameters in a real scenario. For instance, a group of hospitals may have 20 roles, 100 resources, one watchdog, and ten time periods. 
We conclude that \emph{as long as the world state fits in memory, which is the case in these experiments, throughput is not significantly affected by the number of keys or the value size per key}. 

\begin{figure}[t]

 \includegraphics[width=.77\linewidth]{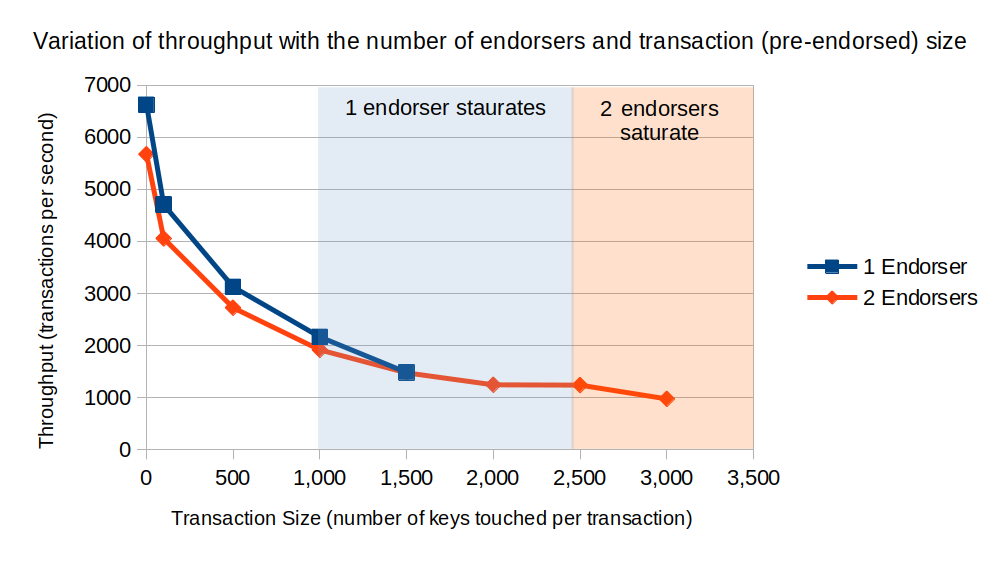}
  \caption{Read throughput (of pre-endorsed transactions) vs.\ transaction size and the number of endorsers (value space = 100; key space = 20,000)}
  \label{exp2}
\end{figure}

\textbf{Effect of transaction size:} Next, we experiment with read transactions touching different numbers of keys: from 1 to 3,000.  Figure~\ref{exp2} shows the results; recall that we report committer throughput, with endorsement time removed (i.e., pre-endorsed transaction throughput), which helps us to zoom in on committer performance. Here we start with a single endorser. We observe that a single endorser cannot keep up when each transaction touches close to 1000 keys (more keys per transaction means more accesses to the world state per transaction during endorsement). At this point, to scale out the system, we add another endorser (Note that here the endorsement policy is one-out-of-two so that the load is distributed among the endorsers.). Having two endorsers allows us to keep up with transactions touching up to 2500 keys.  However, regardless of the number of endorsers, throughput decreases as transaction size increases since large transactions perform more accesses into the world state during commitment (meaning that now, the committer becomes the bottleneck, not the endorsers).  Furthermore, for small transaction sizes, adding a second endorser does not increase (and even slightly decreases) throughput.  We believe that the slight decrease is due to experimental noise as there is no systematic problem with scaling out the endorsers.   

\subsection{IWS vs.\ RWS Design}

\begin{table}[t]
    \centering 
    \footnotesize
        \caption{Comparison of world state designs}
    \label{comp}

    \begin{tabular}{| c | c | c | c | c | c |c |c|}
    \hline
    \multicolumn{2}{| c|}{\textbf{}} & \multicolumn{3}{c|} {\textbf{RWS} (as suggested in~\cite{dubovitskaya2017secure})} & \multicolumn{3}{c|}{\textbf{IWS (\emph{Consentio})}} \\
    \hline
       \#\textbf{R} & \#\textbf{I} &\textbf{\#Keys} & \textbf{\#Hits} & \textbf{TPS} &\textbf{\#Keys }& \textbf{\#Hits }& \textbf{TPS}\\
        \hline
           200 & 200    &  200 &  201 & 1,725 & 200 & 2 & \textbf{6,559}   \\
    \hline
    200 &20K   &   20K &  20K & overload & 200 &  2 &\textbf{6,559}   \\
    \hline
        20K&  200   &   200 &  201 & 1,725 & 20K &  2 & \textbf{6,622}   \\
    \hline
 
    20K & 20K   &   20K &  20K & overload & 20K & 2 & \textbf{6,622}  \\
    \hline

    \multicolumn{8}{|l|}{\footnotesize{\textbf{TPS}=Transactions Per Second; \textbf{R} Resources; \textbf{I} Individuals}}\\
        \multicolumn{8}{|l|}{\footnotesize{\#\textbf{Hits} -  number of keys touched per  transaction}}\\
    \hline
    \end{tabular}

\end{table}

We now compare the RWS world state design suggested in~\cite{dubovitskaya2017secure} with the IWS world state design we use in \emph{Consentio}. We construct four workloads of access requests.
We vary the number of individuals and resources in the world state.  Table~\ref{comp} shows the results, one row per workload (note that the number of keys and keys touched per transaction in a given workload is different for \emph{RWS} and \emph{IWS}). We conclude that \emph{Consentio} has higher throughput in all tested scenarios.  The throughput of \emph{IWS} remains constant as the workload parameters vary, whereas \emph{RWS} overloads when the number of individuals is large. This is because requests to access data in \emph{RWS} correspond to complex transactions that touch many keys, as explained in Section~\ref{analysis}. Even when the number of resources is large, \emph{IWS} performs better than \emph{RWS}. We also experimented with one-out-of-two and two-out-of-two endorsement polices and did not see a significant drop in throughput.



\section{Related Work}
\label{relatedwork}
  
\begin{table}[t]
    \caption{Review of blockchain-based CMSs}
    \label{tab:prev}
\footnotesize
    \centering
    \begin{tabular}{|p{.1\textwidth}|p{.1\textwidth} | p{.08\textwidth}  | p{.09\textwidth}|}
        \hline
\textbf{Papers} & \textbf{Domain Agnostic} & \textbf{Performance Analysis} & \textbf{Implement- abilty}  \\ 
\hline
\cite{zyskind2015decentralizing}  & \cmark & \xmark &  \cmark   \\
\hline
\cite{maesa2017blockchain,choudhury2018enforcing}  & \cmark & \xmark &  \xmark   \\
\hline
\cite{hashemi2017decentralized} & \cmark & \cmark &  \xmark   \\
\hline
\cite{zhang2018smart,rantos2018advocate, bhaskaran2018double} & \xmark & \xmark &  \cmark  \\
\hline
\cite{dias2018blockchain, azaria2016medrec, genestier2017blockchain,hashemi2016world, dubovitskaya2017secure,yue2016healthcare} & \xmark & \xmark &  \xmark \\
\hline

\cite{liang2017integrating, zhang2018block}  & \xmark & \cmark &  \xmark \\
\hline
\emph{Consentio}& \cmark & \cmark &  \cmark       \\
\hline
    \end{tabular}

\end{table}


Table~\ref{tab:prev} summarizes the previous work on blockchains for consent management along the following three axes, and the remainder of this section discusses the work in detail. 
\begin{itemize}
\item \textbf{Domain Agnostic:} Is the system designed for a single domain/application area or for multiple domains?
\item \textbf{Performance analysis:} Is there a performance study, using realistic workloads, in terms of latency/throughput?
\item \textbf{Implementation Details (Replicability): } Are there enough details (in terms of code/pseudocode) to implement/replicate the proposed solution?
\end{itemize}
To the best of our knowledge, \emph{Consentio} is the only blockchain-based CMS that is domain-agnostic, fully implemented, and demonstrated to perform well.


We start with work based on permissionless blockchains.  Some works include proposals but omit implementation details; e.g., \cite{maesa2017blockchain,choudhury2018enforcing, hashemi2017decentralized}. Zyskind et al.~\cite{zyskind2015decentralizing} provide implementation details but do not perform performance analysis. Others include some implementation details, but for specific domains. Zhang et al.~\cite{zhang2018smart} use Ethereum~\cite{buterin2013ethereum} to provide access control for subject-object pairs in the Internet of Things (IoT). Access control contracts perform access right validation based on predefined policies.   Rantos et al.~\cite{rantos2018advocate} provide a GDPR-compliant, Ethereum based framework for managing data from IoT devices. 
Azaria et al.~\cite{azaria2016medrec} use Ethereum to manage authentication, confidentiality, accountability and data sharing. 
However, they focus on giving patients access to their medical histories rather than general consent management. Works such as \cite{hashemi2016world, yue2016healthcare} propose solutions for particular domains and do not provide performance or implementation details.

Moving on to permissioned blockchains, the authors in~\cite{genestier2017blockchain} discuss a Hyperledger-based CMS, but do not provide any design or implementation details. Bhaskaran et al.~\cite{bhaskaran2018double} also provide a Hyperledger-based solution for data sharing but do not include a performance analysis.
The authors in~\cite{dias2018blockchain} propose to use a consortium blockchain for consent management and access control, with the blockchain serving as a repository for access policies. 
However, their design does not consider the granularity of data access, lacks implementation details, and only consider Electronic Health Records.
Dubovitskaya et al.~\cite{dubovitskaya2017secure} design a system for the medical domain. A membership service defines the roles of the actors (doctors and patients). Finally, Hyperledger Fabric executes smart contracts requesting data access. 
However, they assume that only doctors can access patient data, they do not discuss data granularity to which access is granted, and they do not discuss smart contract details.
Furthermore, as discussed earlier, their proposed world state design (RWS) has lower transaction throughput than our design.
Liang et al.~\cite{liang2017integrating} propose to use Fabric channels between individuals and requesters to share data. Again, alongside \cite{zhang2018block}, they do not provide any implementation details.



\section{Conclusion and Future Work}
\label{future}

We presented \textit{Consentio}, a general consent management system using a permissioned blockchain back end: Hyperledger Fabric. 
Using a blockchain allowed us to eliminate the need for a trusted third party to maintain consent settings and transactions. We showed that our solution can be applied to a variety of use cases (Section~\ref{approach}).
An important feature of our solution is that it does not require any modifications to Fabric.   
To preserve compatibility with Fabric, the main technical challenge we addressed was to ensure high throughput and low latency of consent transactions given Fabric's key-value world state.
We analyzed the space of possible world state designs and showed that an efficient world state for consent management can be implemented using Fabric's key-value store (Section~\ref{implementation}). Finally, 
experimental results showed that \emph{Consentio} can handle as many as 6,000 access request per second running on a modest Fabric cluster (Section~\ref{experiments}). 

We focused on the world state design to process consent transactions (i.e., from the Fabric endorser perspective).  An interesting direction for future work on \emph{Consentio} is to develop an efficient data analytics layer for auditing the blockchain.   


\bibliographystyle{IEEEtran}
\bibliography{refs}




%


\end{document}